\documentclass[useAMS,usenatbib]{mn2e}

\usepackage{graphicx}
\usepackage{bm}
\usepackage{lipsum}
\usepackage{color}
\usepackage{longtable}
\newcommand\dsct{$\delta$~Scuti }

\title[$\delta$~Scuti stars in the Galactic bulge]{Frequency analysis of $\delta$~Scuti stars towards the Galactic bulge}
\author[H. Netzel, P. Pietrukowicz, I. Soszy\'nski, M. Wrona]
{H. Netzel$^{1,2,3}$\thanks{E-mail: henia@netzel.pl},
P. Pietrukowicz$^{4}$,
I. Soszy\'nski$^{4}$,
M. Wrona$^{4}$\\
$^{1}$Nicolaus Copernicus Astronomical Center, Polish Academy of Sciences, Bartycka 18, 00-716 Warszawa, Poland\\
$^{2}$Konkoly Observatory, Research Centre for Astronomy and Earth Sciences, E\"otv\"os Lor\'and Research Network (ELKH),\\ H-1121 Konkoly Thege Mikl\'os \'ut 15-17, Budapest, Hungary\\
$^{3}$MTA CSFK Lend\"ulet Near-Field Cosmology Research Group, H-1121 Konkoly Thege Mikl\'os \'ut 15-17, Budapest, Hungary\\
$^{4}$Astronomical Observatory, University of Warsaw, Al. Ujazdowskie 4, 00-478 Warszawa, Poland\\
}

\begin{document}

\date{Accepted . Received ; in original form }

\pagerange{\pageref{firstpage}--\pageref{lastpage}} \pubyear{2021}

\maketitle

\label{firstpage}

\begin{abstract}
We have performed a frequency analysis of 10,092 $\delta$~Scuti-type stars detected in the fourth phase of the Optical Gravitational Lensing Experiment (OGLE) towards the Galactic bulge, which is the most numerous homogeneous sample of \dsct stars observed so far. The main goal was to search for stars pulsating in at least two radial modes simultaneously. We have found 3083 candidates for such stars, which is the largest set obtained to date. Among them, 2655 stars pulsate in two radial modes, 414 stars pulsate in three radial modes, and 14 stars pulsate in four radial modes at the same time. We report the identification of 221 $\delta$~Scuti stars pulsating in the fundamental mode, first overtone, and third overtone simultaneously. We show the most populated Petersen and Bailey diagrams and discuss statistical properties of the identified frequencies based on this numerous sample. Additionally, we present theoretical predictions of period ratios for $\delta$~Scuti stars pulsating in overtones from the fourth to the seventh.
\end{abstract}


\section{Introduction}\label{sec.intro}

$\delta$~Scuti-type stars are intermediate-mass pulsating stars located in the Hertzsprung-Russell (HR) diagram on and close to the main sequence and within the classical instability strip. This location corresponds to spectral types from A0 to F6 and covers luminosity classes from III to V. This diverse group consists of main-sequence stars as well as pre-main-sequence (PMS) and post-main sequence stars. \cite{PMSinst} calculated theoretical instability strip for radial pulsations in PMS stars and concluded that stars with masses above 1.5 M$_\odot$ are situated within the instability strip before they reach the zero-age main sequence (ZAMS). Observationally, first PMS stars of \dsct type were detected by \cite{ngc2264}. In principle, such stars should be rare. However, searches resulted in several discoveries \citep[e.g.][]{diaz}. Even less frequent objects with $\delta$~Scuti-type pulsations should be stars evolving off the main sequence through the Hertzsprung gap. \cite{niu} analyzed photometric and spectroscopic data for AE UMa and concluded that this particular star belongs to Population I \dsct stars evolving through the Hertzsprung gap.

\dsct stars cover a mass range from about 1.0 to 2.5 M$_\odot$ that includes an interesting transition from stars with radiative cores to stars with convective cores. Pulsation periods are shorter than 0.3 d, which is a good criterion to distinguish \dsct stars from nearby groups of pulsating stars such as RR Lyrae-type stars, rapidly oscillating Ap (roAp) stars, and $\gamma$~Doradus-type stars in the HR diagram. However, the instability strip of $\gamma$~Doradus stars overlap with the instability strip of \dsct stars, which results in so-called hybrid stars. In these stars, we observe simultaneous pulsations in pressure modes as well as in gravity modes. For instance, based on time-series multicolor photometry \cite{handler} confirmed the existence of this type of pulsations in object HD~8801.

The majority of \dsct stars pulsate in many non-radial pressure modes excited in the $\kappa$ mechanism. Usually, these modes are of low degree and low order. However, spectroscopic studies confirmed the presence of high-degree modes as well \citep{dsct_spec}. Some \dsct stars pulsate in one or several radial modes with large amplitudes. This subgroup of variables is called High-Amplitude Delta Scuti stars (HADS). The incidence rate of HADS is low. \cite{lee} estimated that this group constitutes only about 0.24 per cent of all Galactic \dsct stars. Several studies also suggest that besides radial pulsations there may be present also low-amplitude non-radial modes. HADS, in contrary to typical \dsct stars, have slow rotation velocities ($v\sin i$ below 30 km/s) which, according to \cite{breger}, is a requirement for high-amplitude radial pulsations. HADS are considered to be the transition objects from large-amplitude Cepheids to small-amplitude pulsating main-sequence stars.

The presence of objects at different evolutionary stage and with different internal structures among \dsct stars makes this group very interesting for asteroseismology and Milky Way studies. Unfortunately, the variety of excited low-amplitude non-radial modes, which are often difficult to identify, limits the usage of \dsct stars. HADS give us an opportunity to derive stellar parameters, because in principle it is easier to identify radial modes based on period ratios. \cite{popielski} discussed the usage of period ratio as a probe of metallicity and applied the idea to double-mode RR Lyrae stars. 

\cite{pietrukowicz} searched for pulsating stars in OGLE-III Galactic disc fields and identified 57 bona fide \dsct stars. Among them, 22 stars pulsate in two radial modes and seven stars pulsate in at least three radial modes. Having at least three radial modes, one can determine crucial stellar parameters such as mass, luminosity, and metallicity. This approach was applied to seven \dsct stars by \cite{pietrukowicz}. In our work, we use the OGLE-IV photometry for over 10,000 recently detected Galactic bulge \dsct pulsators \citep{dsct_collection} to perform a frequency search and to select stars that presumably pulsate in at least two radial modes simultaneously. In Sec.~\ref{sec.analysis}, we describe the dataset and method of the analysis. In Sec.~\ref{sec.results}, we present results of the analysis. In Sec.~\ref{sec.models} we present theoretical predictions for period ratios which are formed by higher overtones. Discussion of our results are presented in Sec.~\ref{sec.discussion}. Sec.~\ref{sec.conslusions} summarizes our findings.

\section{Data and analysis}\label{sec.analysis}

\subsection{Time-series Data}

The Optical Gravitational Lensing Experiment (OGLE) is a long-term variability survey that started in 1992 and is ongoing. In March 2010 it entered its fourth phase with a new mosaic CCD camera installed to the 1.3-m Warsaw telescope located at Las Campanas Observatory, Chile \citep[OGLE-IV,][]{oiv}. OGLE regularly monitors the brightness of about two billion stars towards the Galactic bulge, Galactic disc, and Magellanic Clouds. We used the OGLE-IV photometry for the Galactic bulge fields covering six seasons from 2010 to 2015. We decided to use only those seasons out of all observations carried out in the years 2010--2019 to minimize possible effects of changing period, mean brightness, and amplitude. Firstly, such changes increase the noise level in the power spectrum and would hamper the detection of additional signals. Secondly, during such a long period of observations, proper motions affect the obtained light curves. Removal of the proper motion effect might increase the noise as well. Since in this work, we focus on searches for real additional signals, we did not want to introduce more sources of noise into the data. 

\begin{figure}
\centering
\resizebox{\hsize}{!}{\includegraphics{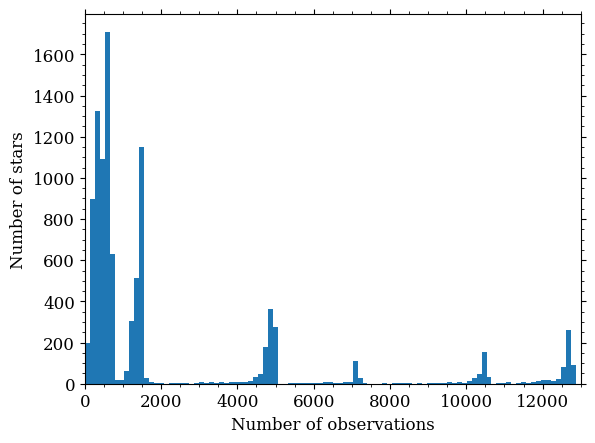}}
\caption{Histogram of the number of $I$-band observations of the Galactic bulge stars.}
\label{fig.hist_points}
\end{figure}

\begin{figure}
\centering
\resizebox{\hsize}{!}{\includegraphics{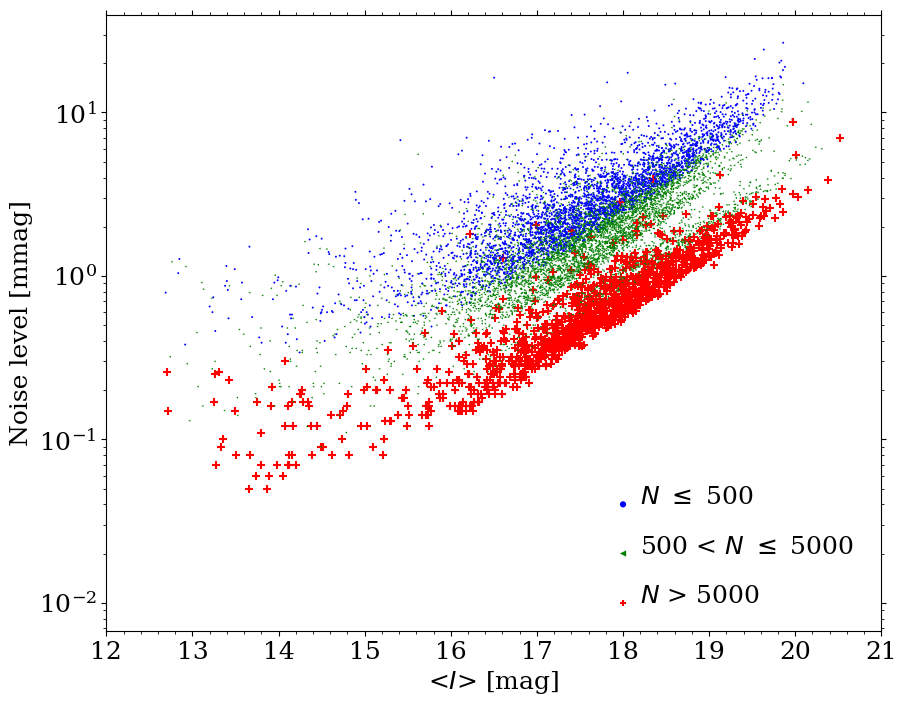}}
\caption{Noise level for \dsct stars in the function of mean $I$-band brightness. \dsct stars are divided into three subgroups depending on the number of observations ($N$) as indicated in the bottom right corner of the figure.}
\label{fig.noise}
\end{figure}

We analyze only $I$-band observations, as these are numerous, hence more suitable for the frequency search than observations in the $V$ band. Our input sample consists of 10,092 \dsct stars published in \cite{dsct_collection}. The original sample was reduced by nineteen sources that doubled other objects. The number of observations for each star varies from field to field. The most frequently observed fields are those located in the densest regions of the Galactic bulge. The largest number of measurements taken over the years 2010--2015 is 12,906, whereas the smallest number is 35. Some fields were observed for a shorter period of time. A histogram of the number of $I$-band measurements is presented in Fig.~\ref{fig.hist_points}. We did not exclude any stars from our analysis based on the number of observations. However, the number of data points affects noise level in the power spectrum. To estimate the noise level for each star, we prewhitened the Fourier spectrum of each star with the dominant frequency, $f$, and its four harmonics and used the mean noise level in the frequency range from $6f$ to $7f$. This relation is presented in Fig.~\ref{fig.noise}. Stars were divided into three subsets based on the number of points: $N\leq 500$, $500<N\leq 5000$, and $N>5000$. For all groups, the noise level increases with the fading brightness. The lowest noise level is for frequently observed stars and the highest noise level is for stars with the smallest number of data points.


\subsection{Frequency Analysis}

The analysis was divided into two steps. The first step was to determine frequencies present in each star. The second step was to analyze the obtained frequencies and to select stars with period ratios falling into known groups of multimode \dsct stars.

The first step was based on Fourier analysis and the consecutive prewhitening method. Due to the large number of stars we used an automatic procedure for this analysis. First, the procedure fitted the dominant frequency and its harmonics to the data in the form of Fourier series:

\begin{equation}
\label{Eq.series}
I(t)=A_0 + \sum A_k \sin(2\pi f_k t + \phi_k),
\end{equation}
where $A_k$, $f_k$, and $\phi_k$ are the amplitude, frequency, and phase of the $k$-th harmonic. Only frequencies fulfilling the criterion $A_k/\sigma_k>4$ were fitted to the data. After prewhitening with the dominant frequency and harmonics, residuals were checked for additional signals in a range 0--80 cycles per day. If any signal in this range was detected, the procedure checked whether its signal-to-noise ratio (S/N) was higher than 4.0. If it was higher, the signal was included in the fit and the program checked again the residuals for the presence of additional frequencies. The procedure stopped once no frequency in the analyzed range fulfilled the criterion $S/N>4.0$ or when the number of independent frequencies exceeded sixteen. Fitting frequencies to the data was done with the help of a non-linear least-squares solver which uses modified Levenberg-Marquardt algorithm. All amplitudes, frequencies, and phases were adjusted while fitting to the data.

We note, that because the OGLE data is ground-based photometry, we had to deal with additional signals that are not connected to the variability of the objects (e.g. daily aliases, instrumental signals). Our procedure took into account the presence of such signals and did not include them in the resultant frequency list. However, we note that the higher priority was to detect signals that form period ratios characteristic for HADS. Therefore, we were not extremely strict with excluding signals and rather checked suspicious signals later during the analysis. In some stars, we observed long-term trends that manifest as signals of low frequencies in the power spectrum. Together with daily aliases, they increase the noise level and hence hamper the detection of low-amplitude additional signals. These trends were removed using a third-order polynomial before the search for additional frequencies.

The program was run on the whole sample of variable stars and provided us a list of frequencies for each star. An example of such list is shown in Tab.~\ref{tab.fit_example}. The automatic procedure provided also a simple interpretation of frequencies, whether they are true independent signals, combination frequencies, or harmonics of signals found previously.

\begin{table*}
\centering
\begin{tabular}{lrrr}
\hline
Signal  &  $f$ [c/d]  & $A$ [mag]  &  $\phi$\\     
  \hline
$f_1$ & 15.972768(5) & 0.048(1) & 5.6(2) \\
$2f_1$ & 31.94554(1) & 0.016(1) & 1.2(4) \\
$f_2$ & 20.56591(2) & 0.012(1) & 5.0(8) \\
$f_2-f_1$ & 4.59314(2) & 0.008(1) & 3.4(8) \\
$f_1+f_2$ & 36.53868(2) & 0.007(1) & 0.4(8) \\
$f_3$ & 46.91555(5) & 0.006(1) & 6(2) \\
$f_4$ & 18.61421(5) & 0.006(1) & 0(2) \\

\end{tabular}
\caption{Example of frequency list generated by the automatic procedure for star OGLE-BLG-DSCT-03647. Consecutive columns provide interpretation of the detected signals, their frequencies, amplitudes, and phases.}
\label{tab.fit_example}
\end{table*}

%

\begin{figure*}
\begin{minipage}{170mm}
\includegraphics[width=\textwidth]{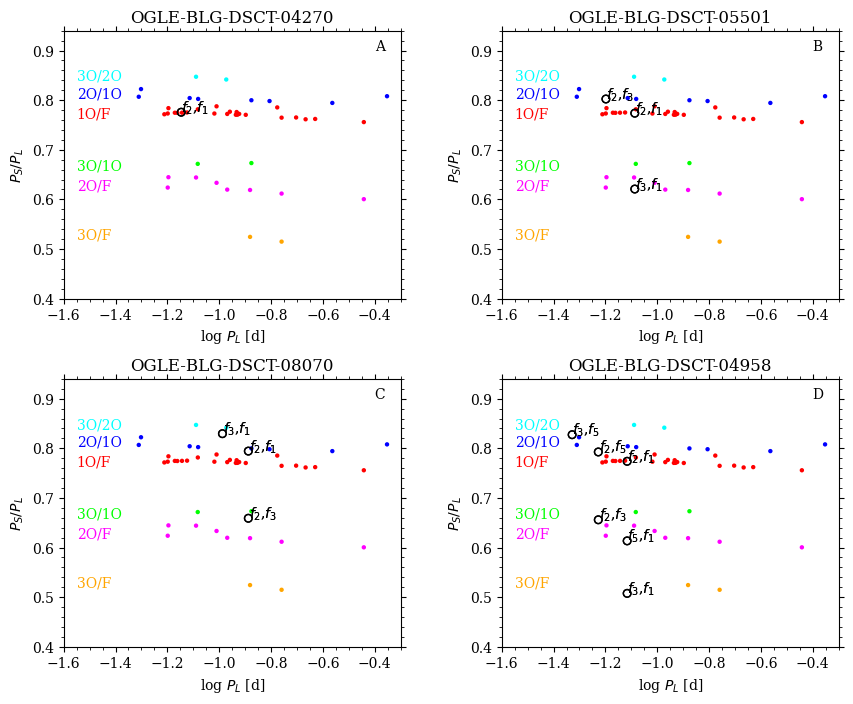}
\caption{Petersen diagrams for selected \dsct stars from our sample. Period ratios formed by detected frequencies are marked with black open circles. Filled circles correspond to multimode \dsct stars from the Galactic disc \citep{pietrukowicz}. On the left side of each sequence it is indicated which modes form the period ratio corresponding to the sequence. Panel A: a double-mode star pulsating in the fundamental mode, $f_1$, and in the first overtone, $f_2$. Panel B: a triple-mode star pulsating in the fundamental mode, $f_1$, in the first overtone, $f_2$, and in the second overtone, $f_3$. Panel C: a triple-mode star pulsating in the first overtone, $f_1$, in the second overtone, $f_2$, and in the third overtone, $f_3$. Panel D: a star pulsating in four modes simultaneously: fundamental mode $f_1$, first overtone $f_2$, second overtone $f_5$, and third overtone $f_3$.}
\label{Fig.pet_examples}
\end{minipage}
\end{figure*}

The second step of the analysis was to select stars for which frequencies fit to the multimode pattern. This was done with the help of the Petersen diagram, which is a diagram of shorter to longer period ratio versus the logarithm of the longer period. Because the main goal of the analysis was to find stars pulsating in radial modes, we narrowed down our sample to stars for which period ratio of two detected periodicities fell into the range 0.75--0.82. In this period ratio, we expect either \dsct stars pulsating in the fundamental mode (F) and first overtone (1O) or the first and second overtone (2O) simultaneously as well as stars with more than two radial modes, two of which form period ratio that falls into this range. This resulted in a sample of 3286 stars. For each star, we plotted all frequencies in the Petersen diagram. For reference, we used stars with identified radial modes from the Galactic disc \citep{pietrukowicz}. All these diagrams were inspected visually. Examples of such diagrams for one double-mode star, two triple-mode stars, and one star with four radial modes are presented in Fig.~\ref{Fig.pet_examples}. Stars were classified based on their position in the Petersen diagram. In some stars, besides frequencies forming the period ratios corresponding to defined sequences for radial pulsations, we detected other signals that do not fit the pattern. Those might be non-radial modes, as they can be present together with radial ones. We assume that all points falling at or close to the sequences are due to radial modes.

Additionally, we checked period ratios corresponding to pulsations F+2O, F+3O, and 1O+3O in order to detect stars that might be candidates for such double-mode radial pulsators. We verified the location of the stars in the Petersen diagram, in the period ratio range corresponding to these multimode pulsations. Then, we selected only objects showing signals being combinations of these modes in the power spectra.

\section{Results}\label{sec.results}

\subsection{Additional signals in \dsct stars}

The majority of the analyzed \dsct stars have more than one periodicity. In Tab.~\ref{Tab.nof}, we list how many stars were found with a various number of frequencies. In the entire sample of 10,092 stars, 2872 objects are single-periodic \dsct pulsators. They can either pulsate in a single radial mode, most likely in the fundamental mode or first overtone, or pulsate in a non-radial mode. Single-periodic \dsct stars are slightly less numerous than stars pulsating in two modes simultaneously. The group of double-mode pulsators is the most abundant in the whole sample and consists of 2947 objects. Although the maximum number of independent frequencies was set to 16, we did not detect any star with this number of signals. Fifteen independent frequencies were detected in two stars, but none of the frequency patterns fits to any radial mode sequence in the Petersen diagram. We did not investigate the nature of additional frequencies except for period ratios corresponding to those of radial modes. Some other detected signals may come from non-radial modes, blending, or other physical phenomena, e.g. eclipses. Signals corresponding to binarity were detected in 14 stars as reported by \cite{dsct_collection}.

\begin{table}
\centering
\begin{tabular}{ccc}
Number of signals & $N_{\rm stars}$ & $N_{\rm stars}/N_{\rm tot}$ \\
\hline
1	&	2872	&	0.28	\%	\\
2	&	2947	&	0.29	\%	\\
3	&	2007	&	0.20	\%	\\
4	&	1177	&	0.12	\%	\\
5	&	565	&	0.06	\%	\\
6	&	267	&	0.03	\%	\\
7	&	116	&	0.01	\%	\\
8	&	54	&	$<$1	\%	\\
9	&	36	&	$<$1	\%	\\
10	&	19	&	$<$1	\%	\\
11	&	14	&	$<$1	\%	\\
12	&	10	&	$<$0.1	\%	\\
13	&	4	&	$<$0.1	\%	\\
14	&	2	&	$<$0.1	\%	\\
15	&	2	&	$<$0.1	\%	\\
16	&	0	&	--	\\
\end{tabular}
\caption{Number of detected independent frequencies in 10,092 analyzed \dsct variables. Consecutive columns provide the following information: the number of frequencies, number of stars, and incidence rate.}
\label{Tab.nof}
\end{table}

\begin{figure}
\centering
\resizebox{\hsize}{!}{\includegraphics{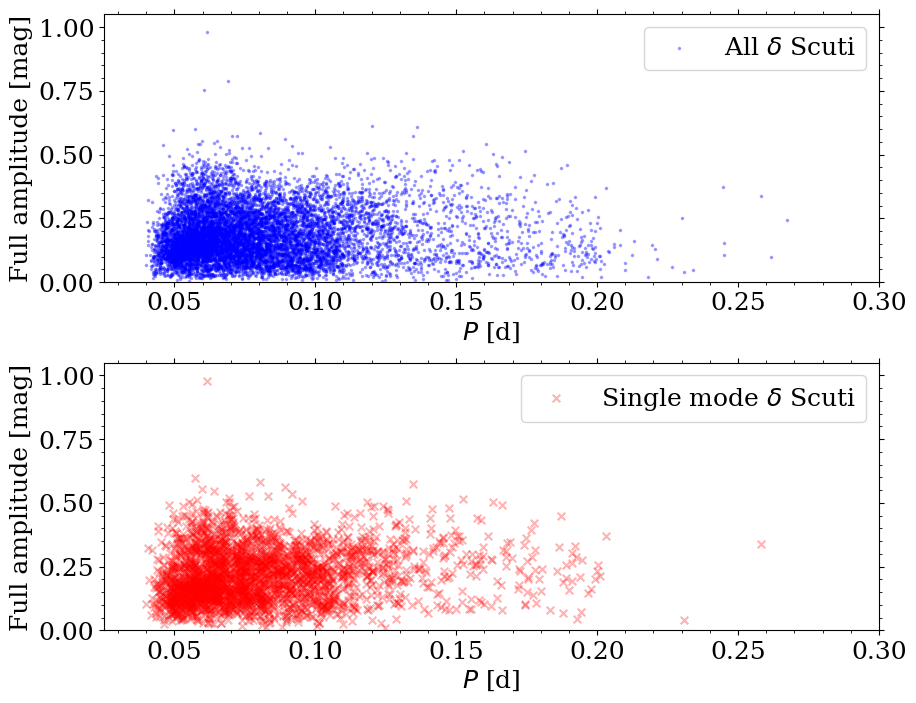}}
\caption{Period--amplitude diagram for all \dsct stars (blue points in the top panel) and single-mode \dsct stars (red crosses in the bottom panel). Stars with very high amplitudes are faint and blended objects.}
\label{fig.single}
\end{figure}

In the top panel of Fig.~\ref{fig.single}, we present a period-amplitude diagram for all analyzed \dsct stars, while in the bottom panel, such a diagram for only single-mode stars. Both groups cover the same range of amplitudes. In the case of single-mode \dsct stars, we observe a slightly shorter range of periods, up to 0.3 d, whereas for all \dsct Scuti stars, there are several objects with periods between 0.3 and 0.35 d. No distinct structures are visible in the diagrams.

\begin{figure}
\centering
\resizebox{\hsize}{!}{\includegraphics{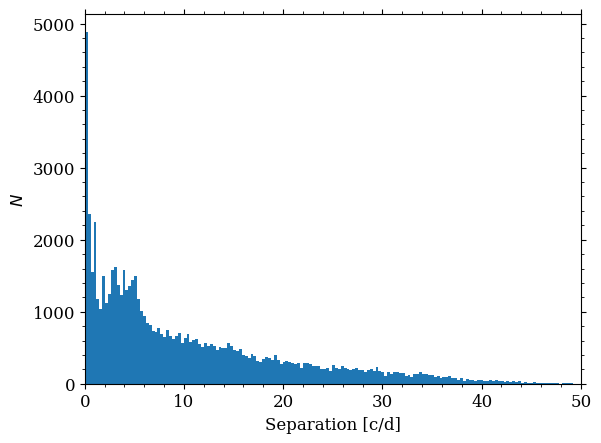}}
\caption{Histogram of frequency separations for all multimode \dsct stars in the sample.}
\label{fig.separacje}
\end{figure}

We were curious whether there are any characteristic equidistant frequency separations between the detected signals in the power spectrum. Specific separations are expected in the case of non-radial pulsations. We calculated separations between frequencies detected in all multimode \dsct stars from our sample. In Fig.~\ref{fig.separacje}, we plot histogram of the obtained separations. For the majority of stars, the separations are very small, but they extend to over 40 c/d. There is no clumping of the values  apart from an excess observed around 3--5 c/d.

\subsection{Multimode radial pulsations in \dsct stars}

\begin{figure}
\centering
\resizebox{\hsize}{!}{\includegraphics{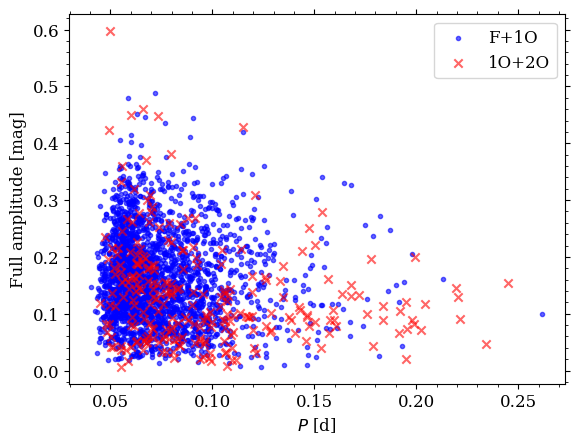}}
\caption{Period-amplitude (Bailey) diagram for \dsct stars classified as objects pulsating in two radial modes simultaneously. With blue points we mark stars pulsating in the fundamental mode and first overtone. Stars classified as objects pulsating in the first and second overtones are marked with red crosses.}
\label{fig.double}
\end{figure}

The main goal of this work was to select candidates for multimode radial pulsators. Results are collected in Tab.~\ref{Tab.class}, where we provide the number of stars together with the incidence rate for various groups of modes. We detected 3083 stars pulsating in at least two radial modes. Among them, 2655 stars are candidates for objects pulsating in two radial modes simultaneously. This corresponds to over 26 per cent of the total sample. The most abundant group is formed of stars pulsating in the fundamental mode and the first overtone (F+1O). This group consists of 2333 objects. The second most numerous group are double-mode stars pulsating in the first and second overtones (1O+2O). Besides the two groups, we found candidates for stars pulsating in the fundamental mode and second overtone (F+2O), fundamental mode and third overtone (F+3O), and first and third overtones (1O+3O). There are 18, 23, and 25 stars in each group, respectively. We note, however, that stars classified as F+2O, F+3O, and 1O+3O pulsators had to fulfill an additional criterion, i.e. the presence of combination frequencies between the radial modes. Hence, the numbers of F+2O, F+3O and 1O+3O stars are the lower limit on the number of these types of double-mode pulsating stars. Four-hundred-and-fourteen \dsct variables are classified as stars pulsating in three radial modes and 14 objects as stars pulsating in four radial modes simultaneously. All stars classified as double-mode radial pulsators are shown in the Bailey diagram in Fig.~\ref{fig.double}, where we plotted period and amplitude of the dominant variability. With different symbols, we mark F+1O and 1O+2O stars. Stars from both groups cover a similar range of periods. Except for one 1O+2O pulsator, of an amplitude of 0.6 mag, the remaining F+1O and 1O+2O pulsators have amplitudes below 0.5 mag. The amplitudes are lower for stars with longer pulsation periods for both groups.

\begin{table}
\centering
\begin{tabular}{lcc}
Modes & $N_{\rm stars}$ & $N_{\rm stars}/N_{\rm tot}$\\
\hline
      F+1O    & 2333 & 23.1 \% \\
     1O+2O    &  256 &  2.5 \% \\
   F+1O+3O    &  221 &  2.2 \% \\
   F+1O+2O    &  145 &  1.4 \% \\
  1O+2O+3O    &   48 &  0.5 \% \\
     1O+3O    &   25 &  0.2 \% \\
      F+3O    &   23 &  0.2 \% \\
      F+2O    &   18 &  0.2 \% \\
F+1O+2O+3O    &   14 &  0.1 \% \\
\end{tabular}
\caption{Census of stars pulsating in at least two radial modes. Consecutive columns provide the type of multiple radial pulsations, number of stars, and fraction of all stars falling into each group.}
\label{Tab.class}
\end{table}

As mentioned above, in the analyzed sample, 414 stars pulsate in three radial modes simultaneously. We classified 145 stars as objects pulsating in the fundamental mode, first, and second overtones (F+1O+2O), 221 stars as objects pulsating in the fundamental mode, first, and third overtones (F+1O+3O), and 48 stars as objects pulsating in the first, second, and third overtones (1O+2O+3O). This corresponds to incidence rates of 1.4, 2.2, and 0.5 per cent for F+1O+2O, F+1O+3O, and 1O+2O+3O pulsators, respectively. Interestingly, the most common type of triple-mode radial pulsations is F+1O+3O. This mode combination was not detected in \dsct stars from the OGLE-III Galactic disc fields. Very recently, \cite{triple_hads} reported the detection of a \dsct variable pulsating in three radial modes corresponding to the fundamental mode, first, and third overtones. This star was detected in the data from the {\it Kepler} space mission, and so far was the only candidate for such type of radial pulsations. Here, we report additional 221 candidates for triple-mode F+1O+3O stars. An interesting group of objects, however, are those pulsating in four radial modes simultaneously. We found 14 candidates for stars pulsating in the fundamental mode, first, second, and third overtones. Tab.~\ref{Tab.4modes} lists periods of the signals identified as radial modes for all the quadruple-mode \dsct pulsators. Analogous information on triple-mode stars is provided in Tab.~A1, sample of which is provided in Tab.~\ref{Tab.3modes}.


\begin{table}
\centering
\begin{tabular}{lcccc}
OGLE-BLG- & $P_{\rm F}$\thinspace [d] & $P_{\rm 1O}$\thinspace [d] & $P_{\rm 2O}$\thinspace [d] & $P_{\rm 3O}$\thinspace [d]    \\
-DSCT-    &                 &                  &                  & \\
\hline
00751     &    0.054775     &     0.042734     &     0.034395     &    0.028151 \\
01345     &    0.054168     &     0.041257     &     0.032238     &    0.026991 \\
03522     &    0.074613     &     0.058328     &     0.047497     &    0.039045 \\
03652     &    0.144371     &     0.111366     &     0.088956     &    0.071905 \\
04220     &    0.134911     &     0.106412     &     0.082682     &    0.072704 \\
04712     &    0.106594     &     0.082204     &     0.064173     &    0.054827 \\
04958     &    0.076545     &     0.059235     &     0.046955     &    0.038854 \\
05749     &    0.161760     &     0.122848     &     0.094629     &    0.082316 \\
06799     &    0.072748     &     0.056377     &     0.045357     &    0.038473 \\
07380     &    0.075060     &     0.058121     &     0.046853     &    0.039223 \\
07587     &    0.071205     &     0.055139     &     0.045207     &    0.037168 \\
07793     &    0.070645     &     0.054875     &     0.042816     &    0.036100 \\
08124     &    0.050560     &     0.038931     &     0.031649     &    0.026535 \\
08774     &    0.116664     &     0.08935      &     0.072157     &    0.061194 \\
\end{tabular}
\caption{\dsct stars pulsating in four radial modes simultaneously. Consecutive columns provide the name, period of the fundamental mode, period of the first overtone up to the period of the third overtone.}
\label{Tab.4modes}
\end{table}

\begin{table}
\centering
\begin{tabular}{lcccc}
OGLE-BLG- & $P_{\rm F}$\thinspace [d] & $P_{\rm 1O}$\thinspace [d] & $P_{\rm 2O}$\thinspace [d] & $P_{\rm 3O}$\thinspace [d]    \\
-DSCT-    &                 &                  &                  & \\
\hline
00011	&	--	&	0.05727	&	0.04661	&	0.03945	\\
00018	&	0.07775	&	0.06058	&	0.04904	&	--	\\
00075	&	0.10713	&	0.08376	&	0.06761	&	--	\\
00078	&	0.15776	&	0.11854	&	--	&	0.07758	\\
00121	&	0.06908	&	0.05344	&	--	&	0.03749	\\
00136	&	0.09938	&	0.07705	&	--	&	0.05021	\\
00143	&	0.07965	&	0.06176	&	--	&	0.04026	\\
00181	&	0.07628	&	0.06013	&	0.04782	&	--	\\
00193	&	0.06612	&	0.05110	&	--	&	0.03562	\\
00336	&	--	&	0.15557	&	0.11981	&	0.09938	\\
\vdots & \vdots & \vdots & \vdots & \vdots \\
\end{tabular}
\caption{\dsct stars pulsating in three radial modes simultaneously. Consecutive columns provide star's ID and periods of the fundamental mode, first, second and third overtones. Mode that is not present is omitted with `--'. Full table is provided in the Appendix available online.}
\label{Tab.3modes}
\end{table}

\begin{figure}
\centering
\resizebox{\hsize}{!}{\includegraphics{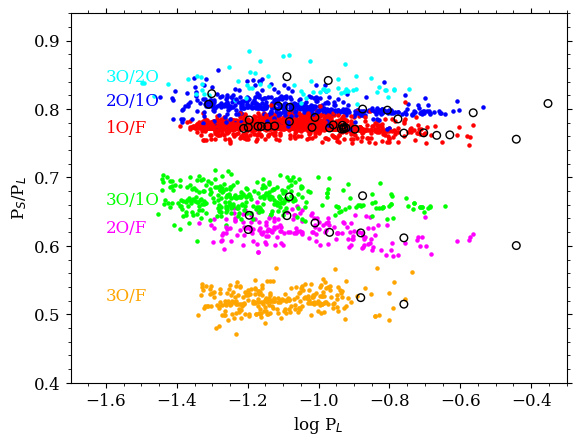}}
\caption{Petersen diagram for \dsct stars pulsating in at least two radial modes simultaneously. Stars from the Galactic bulge sample are marked with filled circles. Black open circles represent stars detected in the Galactic disc by Pietrukowicz et al. (2013).}
\label{fig.pet_rad}
\end{figure}

\begin{figure}
\centering
\resizebox{\hsize}{!}{\includegraphics{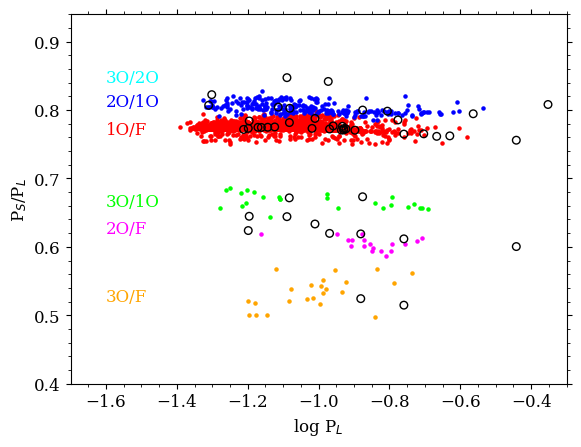}}
\caption{Petersen diagram for \dsct stars pulsating in two radial modes simultaneously. Stars from the bulge sample are marked with filled circles. Black open circles represent disc stars reported in Pietrukowicz et al. (2013).}
\label{fig.pet_double}
\end{figure}

Galactic bulge \dsct stars pulsating in two or more radial modes are plotted in the Petersen diagram in Fig.~\ref{fig.pet_rad} together with Galactic disc \dsct stars analyzed in \cite{pietrukowicz}. Colors indicate period ratios corresponding to various combinations of radial modes. A diagram of solely double-mode radial pulsators is shown in Fig.~\ref{fig.pet_double}. In both figures, the scatter for bulge stars seems to be larger than for disc stars. This can be explained by two facts. First, the total number of disc stars is much smaller than bulge stars. For example, there are only two stars in the sequence corresponding to the period ratio of 3O/F. The same problem is for the sequence corresponding to the period ratio of 3O/1O. Second, \dsct stars observed towards the Galactic bulge represent various populations, hence they have a wide metallicity range. Metal content has a significant influence on the observed period ratios. This is the reason why we did not exclude stars with relatively small departures from the sequences defined by the Galactic disc stars.

\subsection{Light curve decomposion of F+1O pulsators}

\cite{dsct_collection} discussed extensively light curve shapes of \dsct stars based on Fourier coefficients. They studied only single-mode pulsators. Here, we analyze light curve shapes of F+1O pulsators. Thanks to the characteristic period ratio we were able to indicate which mode is the fundamental one and which one is the first overtone. We disentangled the light curves by removing one of the modes. Then, we fitted two separate light curves with a sine series of the same length in the form described with Eq.~\ref{Eq.series}. Amplitude ratios are defined as

\[R_{k1}=\frac{A_k}{A_1}\]

and phase differences are defined as

\[\phi_{k1}=\phi_k-k\phi_1.\]

\begin{figure*}
\begin{minipage}{170mm}
\centering
\resizebox{\hsize}{!}{\includegraphics{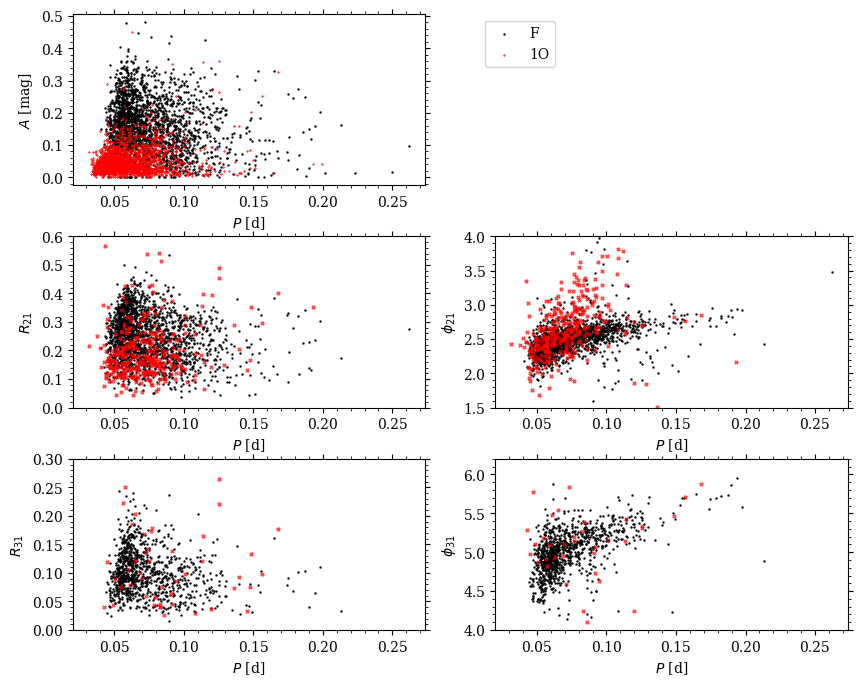}}
\caption{Fourier parameters of disentangled light curves of the fundamental-mode (in black) and first-overtone components (in red) in F+1O stars. Top panel shows peak-to-peak amplitude in the function of period. Middle and bottom panels show the amplitude ratios and phase differences in the function of period.}
\label{fig.fourier}
\end{minipage}
\end{figure*}

In Fig.~\ref{fig.fourier}, we plot peak-to-peak amplitudes and Fourier coefficients in the function of the period for disentangled light curves of F+1O pulsators. Even though for describing light curve shapes we used sine series of the same length, we plotted only those coefficients that are made by significant terms of the sine series ( $A_k/\sigma_k>4$). As expected, the first-overtone components have significantly smaller amplitudes. Except for a few stars, they generally cover half of the range represented by the fundamental mode components. The amplitude ratios $R_{21}$ are on average smaller for the first overtone than for the fundamental mode. Phase differences $\phi_{21}$ and $\phi_{31}$ for fundamental-mode components form clear sequences. There are not many points in the case of $\phi_{31}$ for the first overtone. This is because light curves of the first-overtone components can be described with a shorter Fourier series. In the case of $\phi_{21}$, we observe a slightly larger scatter. Based on the Fourier coefficients alone it is impossible to identify the pulsation modes unambiguously.

\section{Predicted period ratios}\label{sec.models}

\begin{figure*}
\begin{minipage}{170mm}
\includegraphics[width=\textwidth]{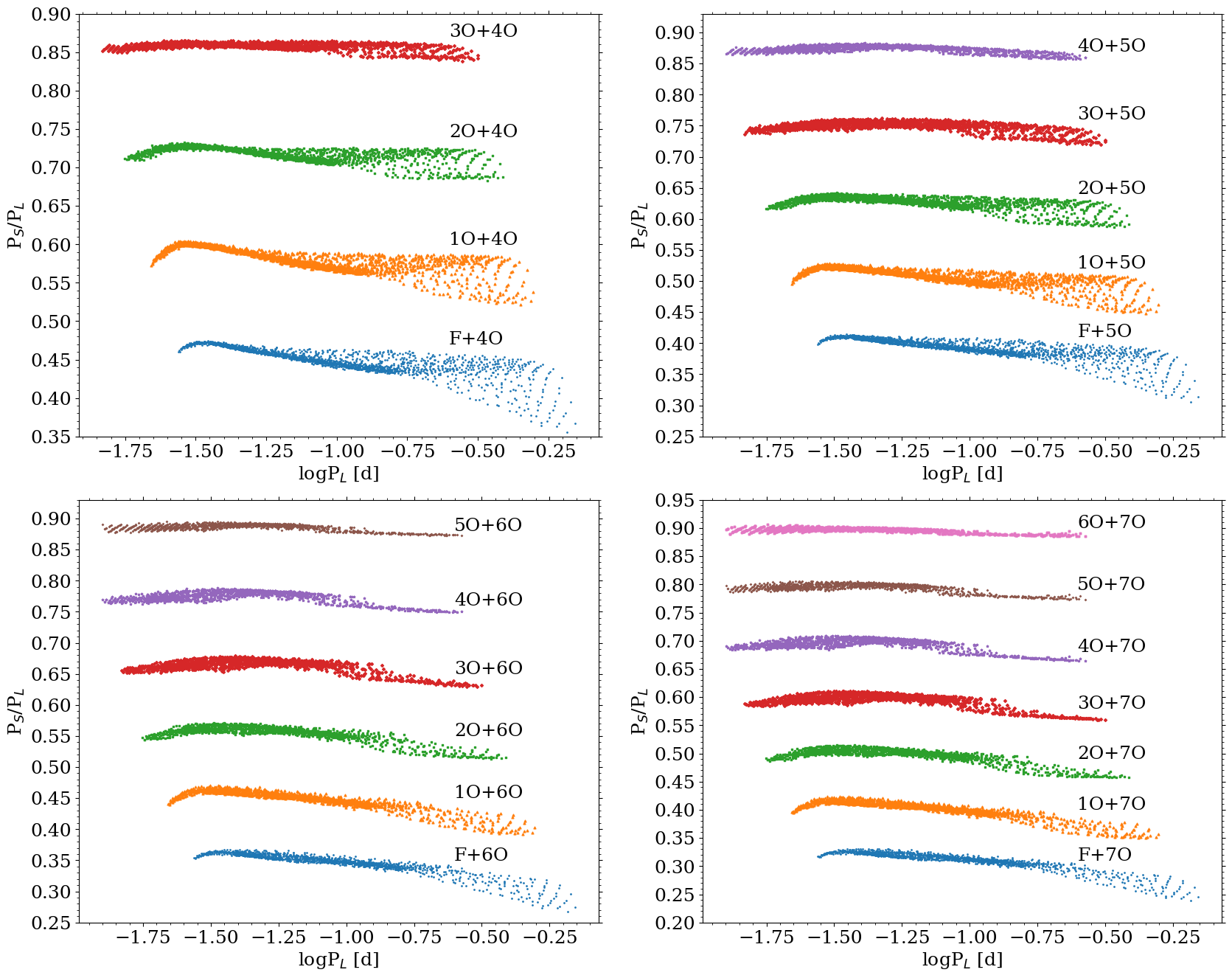}
\caption{Theoretically predicted period ratios for radial modes in \dsct stars.}
\label{Fig.modele}
\end{minipage}
\end{figure*}

We did not find any good candidates for stars pulsating in overtones higher than 3O. During the classification, we, however, compared the results of the frequency analysis with theoretically predicted period ratios for the fourth (4O), fifth (5O), sixth (6O), and seventh overtones (7O). To obtain approximated period ratios we calculated a grid of models for a range of masses and metallicities using the Warsaw evolution and pulsation codes \citep{paczynski,dziembowski1977}. We generated models for masses from 1.0 to 2.5 M$_\odot$ with a step of 0.1 M$_\odot$ and metal content $log~Z$ from $-3.4$ to $-1.1$ with a step of 0.1. We did not include rotation and we set the mixing length parameter to $\alpha_{\rm MLT}=1.5$. We selected models in which certain radial modes are unstable: the fundamental mode, fourth, fifth, sixth, or seventh overtone. Due to the frozen-in description of convection, the code did not allow us to determine the position of the red edge of the instability strip but was able to approximately predict the position of the blue edge. The position of the blue edge also changes due to the frozen-in convection approximation, but only slightly. To limit the number of models, we used the predicted position of the red edge based on observations \citep{alosza_re}. We plot the predicted period ratios of unstable radial modes in Fig.~\ref{Fig.modele}.

\section{Discussion}\label{sec.discussion}

In this work, we analyzed a large sample of 10,092 \dsct stars towards the Galactic bulge from the OGLE-IV survey. Two decades earlier, a catalog of \dsct stars from different sources was prepared by \cite{rodrigez} and contained 636 objects. Unfortunately, the authors did not provide information on the number of HADS and multiperiodicity of stars. \cite{kepler} analyzed 1568 \dsct stars from the {\it Kepler} mission. They focused on the mode content and comparison between the theory and observations. The authors did not provide the number of stars classified as HADS, but the number of stars with one dominant signal. Thanks to excellent quality of the {\it Kepler} photometry, the stars are classified as objects showing one signal with significantly higher amplitude than others, not necessarily as those having only one signal in the power spectrum. \cite{kepler} identified 160 stars with one dominant frequency in the power spectrum, which corresponds to about 10 per cent of their sample. In our sample, we identified 28 per cent of single-periodic stars. The difference stems from a significantly higher noise level in ground-based photometry.

\cite{oiii-lmc} detected 2786 \dsct stars in the OGLE-III data from the Large Magellanic Cloud (LMC). In their sample, they found 92 multimode pulsators. Sixty-seven stars were classified as F+1O pulsators, three stars as 1O+2O pulsators, three stars as 2O+3O pulsators, and three stars as 1O+3O pulsators. There are two triple-mode pulsators. Fourteen objects did not fit the radial mode pattern in the Petersen diagram. This gives an incident rate for F+1O pulsators of 2.4 per cent, which is by an order of magnitude lower than inferred from our analysis. In total, \cite{oiii-lmc} classified 78 objects as multimode radially pulsating \dsct stars. This corresponds to a rate of 3 per cent. It is significantly smaller than the 31 per cent obtained for our sample. This might be due to the difference in observed brightness of \dsct stars located in the LMC fields and Galactic bulge fields. The noise level is higher in more distant LMC objects.

\cite{pietrukowicz} identified 57 \dsct stars in the OGLE-III Galactic disc fields. They detected a single frequency in 28 stars from the analyzed 57 stars, which corresponds to about 49 per cent. This is higher than the 28 per cent obtained here. The incidence rate of stars with two radial modes was 39 per cent and stars with three radial modes was 12 per cent for the disc stars. The incidence rate of double-mode radial pulsations is higher than 26 per cent in the sample analyzed here. In the Galactic bulge sample, we detected triple-mode radial pulsations in 4 per cent of stars, which is three times smaller than in the case of the disc sample. Only one star was identified as a quadruple-mode pulsator in the Galactic disc. This corresponds to the incidence rate of about 2 per cent. The rate of such stars is significantly lower in our analysis (below 1 per cent), but the detected number of stars is 14. Hence, our analysis significantly increased the total number of known stars with four radial modes.

\begin{figure}
\centering
\resizebox{\hsize}{!}{\includegraphics{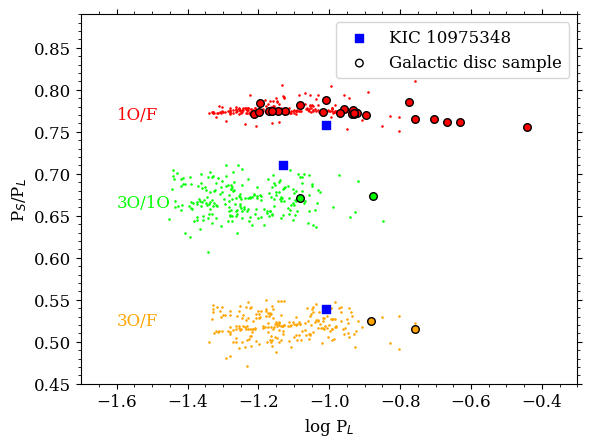}}
\caption{Petersen diagram for \dsct stars classified as F+1O+3O. Red, lime, and orange points correspond to the period ratios of 1O/F, 3O/1O, and 3O/F, respectively. Black open circles represent radially pulsating \dsct stars detected in the Galactic disc \citep{pietrukowicz} forming those period ratios. Blue squares correspond to object KIC 10975348.}
\label{fig.pet_kic}
\end{figure}

We detected a group of stars pulsating in three radial modes F+1O+3O simultaneously. Interestingly, the second overtone is not excited in these stars or has an amplitude below the detection limit. A candidate for such star was reported only recently by \cite{triple_hads}, who analyzed {\it Kepler} photometry for star KIC~10975348. In this object, they detected three independent signals. In KIC~10975348 two of the signals form a period ratio around 0.758, which is close to the sequence corresponding to F and 1O pulsations. It is slightly lower than typical period ratios of 1O/F which implies a high metal content for this star. The third detected signal forms a period ratio of 0.539 with the fundamental mode and a period ratio of 0.71 with the first overtone. These period ratios are close to 3O/F and 3O/1O sequences. In Fig.~\ref{fig.pet_kic}, we show F+1O+3O \dsct pulsators from the OGLE-IV Galactic bulge sample together with variables from the OGLE-III Galactic disc sample \citep{pietrukowicz} and KIC~10975348 \citep{triple_hads}. In the Galactic disc sample, there are only two stars in the 3O/1O and 3O/F sequences. Bulge stars selected in our work form sequences of significant width compared to the sequences formed by the Galactic disc sample. The period ratios of KIC~10975348 are within the scatter observed for the bulge stars. However, two observed period ratios of 3O/1O and 3O/F are higher than the typical values and one period ratio of 1O/F is lower than the typical values. Similarly, the period ratios of KIC~10975348 differ from the sequences defined by the Galactic disc sample. The period ratio of 1O/F is smaller, whereas the period ratios of 3O/1O or 3O/F are higher than observed for the Galactic disc stars. The effect of metallicity should shift all the observed period ratios in the same direction from the main trends. Therefore, the observed period ratios seem inconsistent with the proposed interpretation and it remains uncertain whether KIC~10975348 is a truly F+1O+3O pulsator.


\section{Conclusions}\label{sec.conslusions}

We analyzed the sample of 10,092 \dsct stars observed towards the Galactic bulge during the fourth phase of the OGLE project \citep{dsct_collection}, which is the largest sample of \dsct stars known so far. We searched for multi-periodic pulsators with period ratios characteristic for radial modes. In 2872 stars, we detected only one period. We classified 3083 objects as candidates for stars pulsating in at least two radial modes simultaneously. Among them, 2655 stars were classified as objects pulsating in two radial modes. The most common type of pulsations is the fundamental mode and first overtone, which was detected in 23.1 per cent of the analyzed sample. Another 256 stars were classified as objects pulsating in the first and second overtones. Additionally, we found 25, 18, and 23 stars, which are candidates for 1O+3O, F+2O, and F+3O pulsators, respectively. Another 414 objects were classified as candidates for stars pulsating in three radial modes. In 145 stars, we detected signals that correspond to the fundamental mode, first and second overtones. Forty-eight stars were classified as objects pulsating in the first, second, and third overtones. We recognized 221 stars which are candidates for stars pulsating in the fundamental mode, first overtone, and third overtone, simultaneously. We selected 14 candidates for stars with four radial modes, F+1O+2O+3O. We note here, that besides signals identified as radial modes, in some of the stars we also detected other frequencies that did not fit any radial modes. Some of them are located close to those identified as radial modes. The signals can be either due to additional pulsations in non-radial modes or of other origin. Additional study of selected multi-mode radially pulsating candidates is required to confirm their classification. Such study focused on theoretical modeling of the selected sample is ongoing.

For stars classified as F+1O pulsators, we verified light curve shapes after disentangling signals from the two modes. We did not notice any characteristic clumping in the Fourier space, which would enable unambiguous mode identification based on Fourier decomposition in future studies. We prepared theoretical predictions for period ratios corresponding to pulsations in the fourth, fifth, sixth, and seventh overtones. We did not detect any good candidates for such pulsators in the analyzed sample.

Based on the huge sample of OGLE-IV \dsct stars we tried to find any characteristic pulsation properties of this type of variables. We presented Bailey, Petersen, and Fourier diagrams with the best statistics. There are no specific features in the period-amplitude diagrams. We also searched for frequency grouping in the power spectra. There is only a slight excess of frequency separations around 3-5 c/d, which is difficult to interpret. To conclude, \dsct stars seem to form a very diverse group of pulsators without evident properties.

\section*{Acknowledgments}
HN is supported by the Polish Ministry of Science and Higher Education under grant 0192/DIA/2016/45 within the Diamond Grant Program for years 2016--2021 and by the Foundation for Polish Science (FNP). This project has been supported by the Lend\"ulet Program of the Hungarian Academy of Sciences, project No. LP2018-7/2020. This work has also been supported by the National Science Centre, Poland, grant MAESTRO 2016/22/A/ST9/00009 to IS.

\section*{Data availability}
Photometric data used for this study was published as a part of the OGLE project \citep{dsct_collection} and are available at the OGLE On-line Data Archive.



\appendix
\section{Data tables}
\setlength{\LTcapwidth}{\textwidth}

\begin{onecolumn}
\begin{longtable}{p{1.5cm}p{1.2cm}p{1.2cm}p{1.2cm}p{1.2cm}|p{1.5cm}p{1.2cm}p{1.2cm}p{1.2cm}p{1.2cm}}
\caption{$\delta$ Scuti stars classified as pulsating in three radial modes. Consecutive columns provide star's ID and periods of the fundamental mode, first, second and third overtones. Mode that is not present is omitted with `--'.}\\
OGLE-BLG-DSCT- & $P_{\rm F}$\thinspace [d] & $P_{\rm 1O}$\thinspace [d] & $P_{\rm 2O}$\thinspace [d] & $P_{\rm 3O}$\thinspace [d] & OGLE-BLG-DSCT- & $P_{\rm F}$\thinspace [d] & $P_{\rm 1O}$\thinspace [d] & $P_{\rm 2O}$\thinspace [d] & $P_{\rm 3O}$\thinspace [d]   \\
\hline
\endfirsthead
\multicolumn{5}{l}{\tablename\ \thetable\ -- \textit{Continued from previous page}} \\
OGLE-BLG-DSCT- & $P_{\rm F}$\thinspace [d] & $P_{\rm 1O}$\thinspace [d] & $P_{\rm 2O}$\thinspace [d] & $P_{\rm 3O}$\thinspace [d] & OGLE-BLG-DSCT- & $P_{\rm F}$\thinspace [d] & $P_{\rm 1O}$\thinspace [d] & $P_{\rm 2O}$\thinspace [d] & $P_{\rm 3O}$\thinspace [d]   \\
\hline
\endhead
\hline \multicolumn{5}{l}{\textit{Continued on next page}} \\
\endfoot
\hline
\endlastfoot
00011	&	--	&	0.05727	&	0.04661	&	0.03945	&	05730	&	0.07188	&	0.05563	&	0.04485	&	--	\\
00018	&	0.07775	&	0.06058	&	0.04904	&	--	&	05765	&	0.07040	&	0.05469	&	--	&	0.03669	\\
00075	&	0.10713	&	0.08376	&	0.06761	&	--	&	05797	&	0.07310	&	0.05662	&	--	&	0.03886	\\
00078	&	0.15776	&	0.11854	&	--	&	0.07758	&	05804	&	0.06494	&	0.05027	&	0.04015	&	--	\\
00121	&	0.06908	&	0.05344	&	--	&	0.03749	&	05809	&	--	&	0.07316	&	0.05810	&	0.04728	\\
00136	&	0.09938	&	0.07705	&	--	&	0.05021	&	05813	&	0.09405	&	0.07289	&	--	&	0.04900	\\
00143	&	0.07965	&	0.06176	&	--	&	0.04026	&	05836	&	0.24507	&	0.18685	&	0.14868	&	--	\\
00181	&	0.07628	&	0.06013	&	0.04782	&	--	&	05855	&	0.11330	&	0.08729	&	0.07008	&	--	\\
00193	&	0.06612	&	0.05110	&	--	&	0.03562	&	05867	&	0.05254	&	0.04050	&	--	&	0.02818	\\
00336	&	--	&	0.15557	&	0.11981	&	0.09938	&	05869	&	0.10177	&	0.07911	&	--	&	0.05106	\\
00358	&	0.06059	&	0.04685	&	--	&	0.03099	&	05885	&	0.05687	&	0.04411	&	--	&	0.02843	\\
00394	&	0.14252	&	0.11055	&	0.08915	&	--	&	05907	&	0.06509	&	0.05042	&	--	&	0.03377	\\
00434	&	--	&	0.20614	&	0.16369	&	0.13566	&	05937	&	0.11495	&	0.08946	&	--	&	0.06153	\\
00648	&	0.06346	&	0.04906	&	0.03972	&	--	&	06029	&	0.08555	&	0.06613	&	0.05524	&	--	\\
00758	&	0.08913	&	0.06973	&	--	&	0.04800	&	06039	&	0.05964	&	0.04612	&	--	&	0.03074	\\
00830	&	0.10817	&	0.08351	&	--	&	0.05675	&	06079	&	0.06112	&	0.04717	&	--	&	0.03206	\\
00871	&	0.06834	&	0.05328	&	--	&	0.03644	&	06089	&	0.10008	&	0.07756	&	0.06253	&	--	\\
00880	&	0.06209	&	0.04802	&	0.04049	&	--	&	06091	&	0.08801	&	0.06877	&	--	&	0.04483	\\
00899	&	0.10475	&	0.08104	&	--	&	0.05428	&	06099	&	0.06772	&	0.05269	&	--	&	0.03590	\\
00906	&	--	&	0.19362	&	0.15366	&	0.12716	&	06122	&	0.05885	&	0.04551	&	0.03625	&	--	\\
00922	&	0.10210	&	0.07944	&	--	&	0.05380	&	06154	&	0.05466	&	0.04238	&	0.03491	&	--	\\
00929	&	0.05251	&	0.04086	&	--	&	0.02688	&	06210	&	0.09311	&	0.07395	&	--	&	0.04812	\\
00953	&	0.06857	&	0.05302	&	--	&	0.03550	&	06211	&	0.07828	&	0.06047	&	0.04862	&	--	\\
00957	&	0.06735	&	0.05140	&	0.03982	&	--	&	06281	&	0.07145	&	0.05619	&	--	&	0.03705	\\
00980	&	0.06786	&	0.05274	&	0.04309	&	--	&	06293	&	0.10367	&	0.08017	&	--	&	0.05359	\\
01013	&	0.09848	&	0.07630	&	--	&	0.05229	&	06309	&	0.05787	&	0.04397	&	--	&	0.02858	\\
01050	&	0.05904	&	0.04595	&	--	&	0.03117	&	06323	&	0.10086	&	0.08108	&	--	&	0.05312	\\
01065	&	0.07498	&	0.05799	&	0.04809	&	--	&	06348	&	0.04971	&	0.03845	&	0.03135	&	--	\\
01092	&	0.06326	&	0.04972	&	--	&	0.03373	&	06349	&	0.09000	&	0.07025	&	0.05731	&	--	\\
01119	&	0.12605	&	0.09747	&	0.07842	&	--	&	06369	&	--	&	0.07804	&	0.06264	&	0.05233	\\
01189	&	0.11759	&	0.09126	&	--	&	0.05993	&	06373	&	0.04787	&	0.03676	&	--	&	0.02453	\\
01198	&	--	&	0.14808	&	0.11770	&	0.09712	&	06375	&	0.05569	&	0.04295	&	--	&	0.02877	\\
01223	&	0.09873	&	0.07757	&	--	&	0.05097	&	06396	&	--	&	0.07133	&	0.05581	&	0.04804	\\
01299	&	--	&	0.08599	&	0.06847	&	0.05961	&	06429	&	0.11077	&	0.08570	&	--	&	0.05715	\\
01474	&	0.05758	&	0.04461	&	0.03773	&	--	&	06458	&	0.08837	&	0.06903	&	--	&	0.04490	\\
01534	&	0.05364	&	0.04155	&	0.03323	&	--	&	06474	&	0.06890	&	0.05242	&	--	&	0.03413	\\
01573	&	0.04709	&	0.03640	&	--	&	0.02523	&	06522	&	0.06877	&	0.05318	&	--	&	0.03455	\\
01697	&	0.09457	&	0.07302	&	--	&	0.04878	&	06615	&	0.06345	&	0.04952	&	--	&	0.03156	\\
01778	&	0.11052	&	0.08388	&	0.06928	&	--	&	06624	&	0.04888	&	0.03817	&	--	&	0.02506	\\
01853	&	0.09553	&	0.07466	&	0.06186	&	--	&	06675	&	--	&	0.13158	&	0.10484	&	0.08664	\\
01870	&	0.12981	&	0.10021	&	0.08058	&	--	&	06712	&	0.06177	&	0.04810	&	--	&	0.03184	\\
01893	&	0.10118	&	0.07762	&	--	&	0.05465	&	06748	&	0.07702	&	0.05955	&	0.04926	&	--	\\
01896	&	0.12843	&	0.10240	&	--	&	0.06758	&	06755	&	0.04686	&	0.03624	&	--	&	0.02527	\\
02004	&	0.09091	&	0.07059	&	--	&	0.04759	&	06758	&	0.06008	&	0.04637	&	0.03747	&	--	\\
02034	&	0.06156	&	0.04759	&	0.03788	&	--	&	06774	&	0.04605	&	0.03565	&	0.02914	&	--	\\
02060	&	0.06962	&	0.05435	&	--	&	0.03607	&	06775	&	0.11581	&	0.08965	&	--	&	0.06147	\\
02064	&	0.06121	&	0.04771	&	0.03889	&	--	&	06784	&	0.11692	&	0.09240	&	0.07416	&	--	\\
02109	&	0.06708	&	0.05202	&	0.04457	&	--	&	06798	&	0.08098	&	0.06284	&	--	&	0.04070	\\
02156	&	0.13567	&	0.10455	&	0.08731	&	--	&	06808	&	0.13316	&	0.10270	&	0.08184	&	--	\\
02195	&	0.26725	&	0.20304	&	0.16371	&	--	&	06812	&	0.08976	&	0.06941	&	0.05852	&	--	\\
02486	&	0.08391	&	0.06520	&	0.05489	&	--	&	06840	&	0.10395	&	0.08028	&	0.06450	&	--	\\
02575	&	0.10966	&	0.08485	&	--	&	0.05646	&	06844	&	0.09028	&	0.07047	&	--	&	0.04834	\\
02578	&	--	&	0.14517	&	0.11540	&	0.09507	&	06851	&	0.09580	&	0.07375	&	--	&	0.04951	\\
02605	&	0.06617	&	0.05116	&	--	&	0.03508	&	06896	&	0.05651	&	0.04366	&	--	&	0.02908	\\
02606	&	0.12185	&	0.09389	&	0.07293	&	--	&	06898	&	0.13387	&	0.10307	&	0.08007	&	--	\\
02684	&	0.05794	&	0.04513	&	0.03696	&	--	&	06904	&	0.17478	&	0.14164	&	--	&	0.09121	\\
02686	&	0.07814	&	0.06039	&	--	&	0.03860	&	06933	&	0.10220	&	0.07898	&	--	&	0.05388	\\
02708	&	--	&	0.20608	&	0.16343	&	0.13538	&	06948	&	0.08205	&	0.06360	&	--	&	0.04293	\\
02709	&	0.10624	&	0.08191	&	0.06802	&	--	&	06952	&	0.13768	&	0.10592	&	0.08289	&	--	\\
02713	&	0.07463	&	0.05780	&	--	&	0.03876	&	06956	&	0.09580	&	0.07383	&	--	&	0.05031	\\
02769	&	0.15639	&	0.12003	&	0.09856	&	--	&	06958	&	0.05079	&	0.03928	&	--	&	0.02703	\\
02798	&	0.06850	&	0.05294	&	--	&	0.03474	&	06985	&	--	&	0.06078	&	0.04896	&	0.04015	\\
02812	&	0.05441	&	0.04241	&	--	&	0.02825	&	07039	&	0.05902	&	0.04577	&	--	&	0.03041	\\
02817	&	0.07201	&	0.05581	&	--	&	0.03761	&	07094	&	0.07686	&	0.05984	&	--	&	0.04024	\\
02854	&	0.10973	&	0.08715	&	--	&	0.05591	&	07102	&	0.06770	&	0.05234	&	0.04215	&	--	\\
02867	&	0.12697	&	0.09813	&	0.08195	&	--	&	07160	&	0.04694	&	0.03627	&	--	&	0.02553	\\
02885	&	--	&	0.11563	&	0.09216	&	0.07525	&	07186	&	0.06209	&	0.04800	&	--	&	0.03308	\\
02988	&	--	&	0.22673	&	0.17976	&	0.14909	&	07200	&	0.06801	&	0.05256	&	--	&	0.03416	\\
02989	&	0.07705	&	0.05957	&	--	&	0.03863	&	07241	&	0.07839	&	0.06083	&	0.04914	&	--	\\
03015	&	0.07167	&	0.05534	&	--	&	0.03928	&	07269	&	--	&	0.14140	&	0.11325	&	0.09550	\\
03035	&	0.05668	&	0.04385	&	--	&	0.02994	&	07289	&	0.07248	&	0.05677	&	0.04639	&	--	\\
03046	&	0.09154	&	0.07094	&	--	&	0.04809	&	07300	&	0.11836	&	0.09180	&	--	&	0.05843	\\
03115	&	0.14686	&	0.11352	&	0.09254	&	--	&	07301	&	0.06424	&	0.04981	&	--	&	0.03445	\\
03119	&	0.13806	&	0.10462	&	0.08640	&	--	&	07324	&	0.09674	&	0.07479	&	--	&	0.05114	\\
03153	&	0.10055	&	0.07937	&	0.06378	&	--	&	07345	&	0.07916	&	0.06114	&	0.04761	&	--	\\
03168	&	0.08820	&	0.06998	&	--	&	0.04685	&	07348	&	0.08531	&	0.06576	&	0.05276	&	--	\\
03171	&	0.09488	&	0.07326	&	0.06127	&	--	&	07358	&	0.06729	&	0.05270	&	--	&	0.03591	\\
03183	&	0.08973	&	0.06959	&	--	&	0.04858	&	07362	&	--	&	0.11709	&	0.09490	&	0.07978	\\
03232	&	0.06119	&	0.04742	&	0.03960	&	--	&	07369	&	0.05617	&	0.04360	&	--	&	0.02886	\\
03241	&	0.05991	&	0.04754	&	--	&	0.03236	&	07383	&	0.04927	&	0.03809	&	--	&	0.02644	\\
03261	&	0.14088	&	0.10670	&	--	&	0.07456	&	07388	&	0.11305	&	0.08723	&	0.06995	&	--	\\
03274	&	0.07935	&	0.06128	&	--	&	0.04138	&	07398	&	0.10770	&	0.08372	&	--	&	0.05826	\\
03336	&	0.14052	&	0.10742	&	0.08574	&	--	&	07427	&	0.19371	&	0.14798	&	0.11607	&	--	\\
03373	&	0.06384	&	0.04950	&	0.04010	&	--	&	07449	&	0.11371	&	0.08789	&	--	&	0.06138	\\
03390	&	--	&	0.09956	&	0.07939	&	0.06969	&	07513	&	0.04744	&	0.03662	&	--	&	0.02419	\\
03392	&	--	&	0.19607	&	0.15544	&	0.12863	&	07564	&	--	&	0.15997	&	0.12756	&	0.10810	\\
03399	&	0.05706	&	0.04430	&	--	&	0.02961	&	07589	&	0.06986	&	0.05396	&	0.04338	&	--	\\
03425	&	0.12352	&	0.09430	&	--	&	0.06397	&	07618	&	0.19114	&	0.14612	&	0.11715	&	--	\\
03444	&	0.09624	&	0.07447	&	--	&	0.05085	&	07656	&	0.07766	&	0.06020	&	0.04827	&	--	\\
03473	&	0.07178	&	0.05610	&	0.04406	&	--	&	07669	&	0.10213	&	0.07946	&	--	&	0.05264	\\
03515	&	0.05828	&	0.04536	&	--	&	0.03157	&	07692	&	0.08985	&	0.06933	&	0.05552	&	--	\\
03516	&	--	&	0.19780	&	0.15709	&	0.12836	&	07697	&	0.06460	&	0.05058	&	0.04099	&	--	\\
03553	&	0.05846	&	0.04536	&	--	&	0.02934	&	07714	&	0.08552	&	0.06692	&	--	&	0.04391	\\
03561	&	0.06733	&	0.05224	&	--	&	0.03447	&	07723	&	0.04660	&	0.03607	&	--	&	0.02458	\\
03581	&	--	&	0.11175	&	0.08901	&	0.07349	&	07757	&	0.06951	&	0.05464	&	--	&	0.03583	\\
03598	&	0.08250	&	0.06429	&	--	&	0.04298	&	07775	&	0.08084	&	0.06263	&	--	&	0.04216	\\
03603	&	0.05610	&	0.04343	&	--	&	0.03041	&	07798	&	0.09323	&	0.07234	&	--	&	0.04907	\\
03713	&	0.07120	&	0.05515	&	0.04373	&	--	&	07821	&	0.06219	&	0.04705	&	--	&	0.03283	\\
03749	&	0.10116	&	0.07848	&	0.06079	&	--	&	07833	&	0.12028	&	0.09411	&	0.07554	&	--	\\
03751	&	0.05127	&	0.03911	&	--	&	0.02463	&	07836	&	0.13291	&	0.10089	&	0.08230	&	--	\\
03766	&	0.06597	&	0.05200	&	--	&	0.03411	&	07844	&	0.06462	&	0.05011	&	0.04031	&	--	\\
03771	&	0.13997	&	0.11151	&	0.08993	&	--	&	07868	&	0.07111	&	0.05582	&	--	&	0.03707	\\
03773	&	0.11241	&	0.08472	&	--	&	0.05930	&	07885	&	0.11299	&	0.08742	&	--	&	0.05901	\\
03780	&	0.09660	&	0.07460	&	0.05994	&	--	&	07891	&	0.07023	&	0.05468	&	--	&	0.03550	\\
03815	&	0.05833	&	0.04529	&	--	&	0.03004	&	07918	&	--	&	0.13302	&	0.10597	&	0.08765	\\
03820	&	0.05823	&	0.04495	&	--	&	0.03061	&	07926	&	0.15753	&	0.12092	&	--	&	0.08356	\\
03838	&	0.10761	&	0.08358	&	--	&	0.05758	&	07927	&	0.05753	&	0.04481	&	0.03727	&	--	\\
03843	&	0.06157	&	0.04783	&	0.03863	&	--	&	07952	&	0.06835	&	0.05307	&	--	&	0.03417	\\
03849	&	0.06623	&	0.05174	&	--	&	0.03552	&	07956	&	0.04852	&	0.03749	&	--	&	0.02482	\\
03857	&	0.08536	&	0.06682	&	0.05502	&	--	&	07969	&	0.08779	&	0.06821	&	--	&	0.04470	\\
03862	&	0.09835	&	0.07680	&	0.05970	&	--	&	07980	&	0.06516	&	0.05067	&	--	&	0.03364	\\
03874	&	0.26655	&	0.19914	&	0.16239	&	--	&	07981	&	0.07068	&	0.05474	&	--	&	0.03448	\\
03887	&	0.07403	&	0.05715	&	0.04589	&	--	&	07998	&	0.08542	&	0.06800	&	--	&	0.04462	\\
03891	&	0.10296	&	0.08032	&	0.06498	&	--	&	08009	&	--	&	0.08972	&	0.07222	&	0.06020	\\
03892	&	0.05342	&	0.04147	&	--	&	0.02667	&	08010	&	0.08489	&	0.06727	&	0.05422	&	--	\\
03910	&	--	&	0.15085	&	0.11984	&	0.09895	&	08021	&	0.08225	&	0.06530	&	--	&	0.04175	\\
03943	&	0.07095	&	0.05609	&	--	&	0.03711	&	08050	&	0.16804	&	0.12628	&	0.09867	&	--	\\
03949	&	0.14844	&	0.11417	&	--	&	0.07407	&	08070	&	--	&	0.12953	&	0.10290	&	0.08536	\\
03986	&	0.08931	&	0.06894	&	0.05538	&	--	&	08075	&	0.20811	&	0.15851	&	0.12237	&	--	\\
03990	&	0.06111	&	0.04738	&	--	&	0.03150	&	08132	&	0.04942	&	0.03814	&	--	&	0.02614	\\
04045	&	0.05575	&	0.04341	&	--	&	0.02990	&	08141	&	0.09863	&	0.07692	&	0.06085	&	--	\\
04071	&	0.09699	&	0.07647	&	0.06213	&	--	&	08155	&	0.06917	&	0.05398	&	--	&	0.03727	\\
04075	&	0.05999	&	0.04661	&	--	&	0.03247	&	08165	&	0.07966	&	0.06142	&	0.04926	&	--	\\
04116	&	0.10203	&	0.07906	&	--	&	0.05255	&	08200	&	--	&	0.15760	&	0.12540	&	0.10376	\\
04164	&	0.10324	&	0.07989	&	--	&	0.05200	&	08216	&	0.05847	&	0.04516	&	--	&	0.03034	\\
04210	&	0.10629	&	0.08414	&	0.06755	&	--	&	08235	&	0.06723	&	0.05215	&	--	&	0.03479	\\
04212	&	--	&	0.04763	&	0.03761	&	0.03147	&	08238	&	0.05503	&	0.04261	&	--	&	0.02923	\\
04232	&	--	&	0.12723	&	0.10122	&	0.08347	&	08253	&	0.05605	&	0.04332	&	0.03507	&	--	\\
04237	&	0.08320	&	0.06433	&	0.05165	&	--	&	08263	&	0.07220	&	0.05497	&	--	&	0.03612	\\
04260	&	0.06319	&	0.04930	&	--	&	0.03335	&	08266	&	0.09773	&	0.07621	&	0.06160	&	--	\\
04276	&	0.06804	&	0.05348	&	--	&	0.03632	&	08279	&	0.05033	&	0.03886	&	0.03049	&	--	\\
04278	&	0.09513	&	0.07325	&	--	&	0.04741	&	08319	&	0.07324	&	0.05671	&	0.04555	&	--	\\
04298	&	0.08051	&	0.06234	&	--	&	0.04265	&	08325	&	0.06937	&	0.05430	&	0.04283	&	--	\\
04300	&	0.08951	&	0.06927	&	0.05547	&	--	&	08329	&	0.09968	&	0.07719	&	--	&	0.05183	\\
04311	&	0.06154	&	0.04792	&	--	&	0.03175	&	08335	&	0.11280	&	0.08770	&	--	&	0.05753	\\
04332	&	0.05080	&	0.03931	&	0.03250	&	--	&	08345	&	0.06795	&	0.05273	&	--	&	0.03623	\\
04348	&	0.08745	&	0.06776	&	--	&	0.04400	&	08348	&	0.08998	&	0.06968	&	--	&	0.04607	\\
04349	&	0.05811	&	0.04509	&	0.03720	&	--	&	08362	&	--	&	0.07645	&	0.06219	&	0.05073	\\
04376	&	0.09283	&	0.07261	&	--	&	0.04868	&	08369	&	0.05246	&	0.04059	&	--	&	0.02673	\\
04440	&	0.06041	&	0.04661	&	--	&	0.03115	&	08372	&	--	&	0.08044	&	0.06473	&	0.05377	\\
04458	&	0.08868	&	0.06852	&	0.05511	&	--	&	08377	&	0.12473	&	0.09588	&	0.07669	&	--	\\
04473	&	0.05301	&	0.04097	&	0.03218	&	--	&	08379	&	0.20068	&	0.15055	&	0.12045	&	--	\\
04542	&	0.08691	&	0.06737	&	0.05402	&	--	&	08388	&	0.05971	&	0.04602	&	--	&	0.03028	\\
04545	&	0.05004	&	0.03867	&	--	&	0.02661	&	08408	&	0.07410	&	0.05874	&	--	&	0.03848	\\
04559	&	0.05459	&	0.04235	&	--	&	0.02812	&	08463	&	0.05564	&	0.04321	&	--	&	0.02739	\\
04561	&	0.12311	&	0.09719	&	0.07806	&	--	&	08480	&	0.05516	&	0.04273	&	0.03446	&	--	\\
04568	&	0.08642	&	0.06722	&	--	&	0.04664	&	08534	&	0.05342	&	0.04120	&	0.03291	&	--	\\
04603	&	--	&	0.14171	&	0.11315	&	0.09391	&	08561	&	0.04966	&	0.03846	&	--	&	0.02547	\\
04604	&	0.08640	&	0.06669	&	0.05349	&	--	&	08585	&	0.10638	&	0.08245	&	--	&	0.05743	\\
04643	&	0.08652	&	0.06719	&	0.05422	&	--	&	08588	&	0.06445	&	0.04988	&	0.04153	&	--	\\
04663	&	0.07531	&	0.05813	&	0.04529	&	--	&	08589	&	0.07227	&	0.05659	&	--	&	0.03710	\\
04664	&	0.10469	&	0.08080	&	0.06524	&	--	&	08598	&	0.11178	&	0.08625	&	0.06824	&	--	\\
04683	&	0.07294	&	0.05674	&	--	&	0.04002	&	08632	&	0.07083	&	0.05518	&	0.04461	&	--	\\
04720	&	0.08574	&	0.06666	&	--	&	0.04355	&	08642	&	0.06710	&	0.05074	&	0.04047	&	--	\\
04750	&	0.07231	&	0.05594	&	--	&	0.03761	&	08647	&	0.06388	&	0.04942	&	--	&	0.03379	\\
04777	&	0.05455	&	0.04222	&	--	&	0.02723	&	08671	&	0.12075	&	0.09306	&	--	&	0.06329	\\
04781	&	0.08713	&	0.06757	&	--	&	0.04571	&	08687	&	0.05444	&	0.04230	&	--	&	0.02746	\\
04783	&	--	&	0.15992	&	0.12748	&	0.10574	&	08709	&	0.07300	&	0.05662	&	--	&	0.03668	\\
04796	&	0.05560	&	0.04323	&	--	&	0.02886	&	08748	&	0.07421	&	0.05741	&	--	&	0.03777	\\
04811	&	0.05351	&	0.04146	&	0.03255	&	--	&	08878	&	--	&	0.07977	&	0.06379	&	0.05302	\\
04850	&	0.06728	&	0.05212	&	--	&	0.03514	&	08879	&	0.07907	&	0.06126	&	--	&	0.04135	\\
04886	&	0.07451	&	0.05897	&	--	&	0.04091	&	08920	&	0.05238	&	0.04056	&	--	&	0.02532	\\
04913	&	--	&	0.17671	&	0.14014	&	0.11836	&	08936	&	0.06523	&	0.05044	&	0.04070	&	--	\\
04936	&	0.11774	&	0.09209	&	0.07463	&	--	&	08958	&	0.07793	&	0.06037	&	--	&	0.04056	\\
04947	&	0.08381	&	0.06529	&	--	&	0.04294	&	08984	&	0.05811	&	0.04542	&	0.03696	&	--	\\
04955	&	0.10526	&	0.07978	&	0.06663	&	--	&	08989	&	0.06728	&	0.05342	&	0.04373	&	--	\\
04970	&	0.08629	&	0.06664	&	0.05358	&	--	&	09015	&	--	&	0.07507	&	0.05908	&	0.04964	\\
04971	&	--	&	0.18468	&	0.14630	&	0.12069	&	09031	&	0.10386	&	0.08051	&	--	&	0.05473	\\
04976	&	0.07827	&	0.06056	&	0.04733	&	--	&	09041	&	--	&	0.08383	&	0.06622	&	0.05644	\\
04982	&	0.05676	&	0.04388	&	0.03446	&	--	&	09055	&	0.05471	&	0.04204	&	0.03383	&	--	\\
05003	&	--	&	0.15232	&	0.12108	&	0.10001	&	09058	&	0.08041	&	0.06205	&	--	&	0.04085	\\
05037	&	0.07117	&	0.05505	&	--	&	0.03589	&	09071	&	0.06233	&	0.04816	&	0.03902	&	--	\\
05049	&	0.05262	&	0.04065	&	--	&	0.02797	&	09074	&	0.06019	&	0.04675	&	--	&	0.03048	\\
05056	&	0.13231	&	0.10188	&	0.07904	&	--	&	09080	&	0.06657	&	0.05169	&	--	&	0.03329	\\
05068	&	--	&	0.18621	&	0.14830	&	0.12330	&	09109	&	--	&	0.12151	&	0.09671	&	0.07981	\\
05082	&	0.04567	&	0.03529	&	--	&	0.02281	&	09122	&	--	&	0.08508	&	0.06823	&	0.05703	\\
05115	&	0.09895	&	0.07693	&	0.06344	&	--	&	09153	&	0.10064	&	0.07793	&	--	&	0.05163	\\
05117	&	0.08683	&	0.06793	&	0.05714	&	--	&	09170	&	0.08871	&	0.06905	&	--	&	0.04725	\\
05136	&	0.05944	&	0.04592	&	--	&	0.02994	&	09177	&	--	&	0.14894	&	0.11885	&	0.10295	\\
05190	&	0.07345	&	0.05918	&	--	&	0.03813	&	09199	&	0.07233	&	0.05604	&	--	&	0.03785	\\
05207	&	0.06937	&	0.05383	&	--	&	0.03662	&	09226	&	--	&	0.08207	&	0.06362	&	0.05628	\\
05245	&	0.07664	&	0.05917	&	--	&	0.03973	&	09241	&	0.08137	&	0.06295	&	0.05289	&	--	\\
05260	&	0.05724	&	0.04452	&	--	&	0.03046	&	09260	&	0.08861	&	0.06853	&	--	&	0.04606	\\
05303	&	0.06756	&	0.05161	&	0.03996	&	--	&	09335	&	0.09732	&	0.07646	&	--	&	0.05043	\\
05333	&	--	&	0.13050	&	0.10385	&	0.08604	&	09364	&	0.06525	&	0.05067	&	0.03961	&	--	\\
05356	&	0.06610	&	0.05112	&	--	&	0.03632	&	09371	&	0.09370	&	0.07247	&	--	&	0.05022	\\
05361	&	0.07466	&	0.05904	&	--	&	0.03881	&	09472	&	0.07293	&	0.05649	&	--	&	0.03785	\\
05371	&	0.06851	&	0.05310	&	0.04292	&	--	&	09493	&	0.07490	&	0.05786	&	0.04705	&	--	\\
05401	&	0.09677	&	0.07477	&	--	&	0.04990	&	09501	&	0.06633	&	0.05162	&	0.04244	&	--	\\
05413	&	0.08262	&	0.06400	&	--	&	0.04309	&	09517	&	0.08307	&	0.06437	&	--	&	0.04175	\\
05421	&	0.06303	&	0.04863	&	--	&	0.03380	&	09524	&	0.05173	&	0.04011	&	--	&	0.02655	\\
05428	&	0.07761	&	0.05994	&	--	&	0.04025	&	09566	&	0.07385	&	0.05729	&	--	&	0.03785	\\
05468	&	0.10423	&	0.08080	&	--	&	0.05396	&	09581	&	0.05767	&	0.04487	&	--	&	0.02984	\\
05477	&	0.07137	&	0.05517	&	0.04436	&	--	&	09637	&	0.05263	&	0.04085	&	--	&	0.02723	\\
05480	&	0.08081	&	0.06257	&	--	&	0.04199	&	09640	&	--	&	0.16176	&	0.12848	&	0.10364	\\
05501	&	0.08191	&	0.06337	&	0.05085	&	--	&	09666	&	0.27319	&	0.21228	&	0.16877	&	--	\\
05514	&	0.10416	&	0.08141	&	--	&	0.05698	&	09728	&	--	&	0.10558	&	0.08550	&	0.07145	\\
05548	&	--	&	0.11320	&	0.09035	&	0.07428	&	09746	&	0.09799	&	0.07745	&	0.06154	&	--	\\
05572	&	0.09416	&	0.07267	&	0.05755	&	--	&	09778	&	0.04818	&	0.03722	&	--	&	0.02609	\\
05580	&	0.06215	&	0.04805	&	--	&	0.03112	&	09840	&	0.12257	&	0.09468	&	0.07604	&	--	\\
05589	&	0.19935	&	0.15262	&	0.12808	&	--	&	09888	&	0.08871	&	0.06851	&	0.05747	&	--	\\
05617	&	0.09154	&	0.07221	&	--	&	0.04987	&	09903	&	0.06119	&	0.04721	&	0.03683	&	--	\\
05622	&	0.11379	&	0.08995	&	--	&	0.05747	&	09905	&	0.07090	&	0.05483	&	0.04270	&	--	\\
05638	&	0.05860	&	0.04545	&	--	&	0.02760	&	09913	&	0.05354	&	0.04134	&	--	&	0.02751	\\
05692	&	0.11939	&	0.09206	&	--	&	0.06120	&	10048	&	0.06033	&	0.04658	&	--	&	0.02987	\\
05695	&	0.05903	&	0.04595	&	--	&	0.03229	&	10050	&	0.04966	&	0.03868	&	--	&	0.02690	\\
05720	&	0.12480	&	0.09612	&	0.07621	&	--	&	10100	&	0.09560	&	0.07382	&	0.05925	&	--	\\
\end{longtable}
\end{onecolumn}


\bsp

\label{lastpage}

\end{document}